\documentclass[aps,prl,twocolumn,superscriptaddress, showpacs]{revtex4-1}

\newcommand{\bra}[1]{\left\langle{#1}\right\vert}
\newcommand{\ket}[1]{\left\vert{#1}\right\rangle}
\usepackage{graphicx}
\usepackage{amsmath}

\begin{document}


\title{On the effects of self- and cross-phase modulation on photon purity for four-wave mixing photon-pair sources}


\author{Bryn Bell}
\email[]{bbell@physics.usyd.edu.au}
\affiliation{Department of Electrical and Electronic Engineering, University of Bristol,  Bristol BS8 1UB, UK}
\affiliation{Centre for Ultrahigh bandwidth Devices for Optical Systems (CUDOS), Institute of Photonics and Optical Science (IPOS), School of Physics, University of Sydney, NSW 2006, Australia}

\author{Alex McMillan}
\author{Will McCutcheon}
\author{John Rarity}

\affiliation{Department of Electrical and Electronic Engineering, University of Bristol,  Bristol BS8 1UB, UK}


\date{\today}

\begin{abstract}
We consider the effect of self-phase modulation and cross-phase modulation on the joint spectral amplitude of photon pairs generated by spontaneous four-wave mixing. In particular, the purity of a heralded photon from a pair is considered, in the context of schemes that aim to maximise the purity and minimise correlation in the joint spectral amplitude using birefringent phase-matching and short pump pulses. We find that non-linear phase modulation effects will be detrimental, and will limit the quantum interference visibility that can be achieved at a given generation rate. An approximate expression for the joint spectral amplitude with phase modulation is found by considering the group velocity walk-off between each photon and the pump, but neglecting the group-velocity dispersion at each wavelength. The group-velocity dispersion can also be included with a numerical calculation, and it is shown that it only has a small effect on the purity for the realistic parameters considered.
\end{abstract}

\pacs{42.50.Dv, 03.67.Bg}

\maketitle
\section{Introduction}

Single photon sources are a vital component of developing quantum technologies such as quantum cryptography~\cite{gisin_qcom}, linear optical quantum computing~\cite{ladd_qcpu}, and quantum metrology~\cite{gio_qmet}, and improved sources need to be developed to enable these applications. In addition to requiring high efficiency, on-demand single photons, many applications require that the photons be generated in a single-mode, with well defined spatial characteristics and a fourier transform limited spectral-temporal shape. This allows two separate photons to be indistinguishable and to undergo high quality quantum interference~\cite{hom} - this, in turn, makes possible fundamental operations such as teleportation of the photon~\cite{ma_qtele}, and two photon logic gates~\cite{obrien_cnot}.

Photon pairs generated in a nonlinear medium by spontaneous parametric downconversion (SPDC) or four-wave mixing (FWM) are often used as a source of single photons, with one of the photons detected to give a heralding signal, indicating the presence of the other~\cite{fasel_SPDC, mcmillan_FWM}. Although this method is inherently non-deterministic, through the multiplexing of many such sources and the use of active switching, it is in principle possible to construct a source arbitrarily close to a deterministic source~\cite{ma_multiplex, collins_multiplex}. However, the photons of a pair are generally correlated in frequency or time, which means that the single photons will arrive in a statistical mix of multiple modes. Narrow spectral filtering of the single photons can force them into a single mode, but at a cost to the overall transmission and heralding efficiency, which reduces the usefulness of the source. Possible solutions to this problem have been demonstrated based on consideration of the joint spectral amplitude (JSA) of a pair: with careful choice of wavelengths, or by engineering the dispersion properties of the nonlinear medium, the degree of correlation can be minimised, allowing quantum interference to take place without narrow filtering~\cite{uren_purestate}. For SPDC in bulk crystals interference visibilities as high as $94.5\%$ have been observed without filtering~\cite{mosley_purestate}. For FWM in birefringent optical fibres unfiltered visibilities have tended to fall short of theoretical estimates, often in the range $70 - 80\%$, which is far from sufficient for scalable use in communications or computing~\cite{halder_purestate, cohen_purestate, clark_purestate, soller_purestate}. It has been suggested that inhomogeneity along the length of fibres due to fabrication imperfections is responsible for the short-fall in visibility, and it has been shown theoretically that a large inhomogenity can reduce the visibility~\cite{cui_inhom}. However, it is also expected to create a broadening and modulation in the spectra of the individual photons, which should be easily detected, and in some situations a small amount of inhomogeneity could actually improve the interference visibility~\cite{spring_chipsource}.

Here, we consider the effects of parasitic nonlinear processes on the JSA and the interference visibility, namely self-phase modulation (SPM) and cross-phase modulation (XPM), which are not present in a $\chi^{(2)}$ nonlinear medium such as the crystals used for SPDC, but are potentially significant in a $\chi^{(3)}$ nonlinearity such as fibre~\cite{helt_parasitic}. Previous calculations of the JSA have tended to account for SPM and XPM in a simplistic form which is only exact in a continuous wave (CW) regime, whereas the relevant schemes to make the JSA uncorrelated rely on the use of short pulses. We show that these effects can cause a reduction in interference visibility, especially when a source is operated at a high pair generation rate, beginning from an analytic explanation then progressing to a numerical model taking full account of SPM, XPM, and dispersion.

\section{Joint Spectral Amplitudes}

In pair production through FWM, a bright pump laser is used to power the process. As the pump pulse propagates through a $\chi^{(3)}$ medium, two pump photons may be spontaneously annihilated, with a correlated signal-idler photon pair created. The frequencies of the signal and idler are constrained by the conservation of energy and momentum:
\begin{equation}
\begin{array}{c}
\Delta\omega=2\omega_p -\omega_s-\omega_i=0\\
\Delta\beta=2\beta_p-\beta_s-\beta_i-2\gamma P=0.
\end{array}
\end{equation}
$\omega_{p,s,i}$ refer to the frequency of pump, signal, idler, and $\beta_{p,s,i}$ to the wave vectors. The $2\gamma P$ term arises from SPM and XPM, as the intense pump in the nonlinear medium will slightly modify the wave vectors, with $\gamma$ an effective nonlinear coefficient, and $P$ the peak power of the pump.

These conditions are not exact and the photons will have some bandwidth centered on an exact solution. The finite bandwidth of the pump will introduce some uncertainty to the $\Delta\omega$ condition, and for a fibre of finite length, there is some uncertainty in the phase matching which permits small values of $\Delta\beta$. The JSA can be simply expressed as the product of an energy matching and a phase-matching function~\cite{uren_purestate}:
\begin{equation}
\text{JSA}(\omega_s, \omega_i)=F\times G
\label{JSA_FxG}
\end{equation}
with
\begin{equation}
F=\iint d\omega_{p1}d\omega_{p2}E(\omega_{p1})E(\omega_{p2})\delta(\omega_{p1}+\omega_{p2}-\omega_s-\omega_i)
\end{equation}
\begin{equation}
G=e^{i\Delta\beta L/2}\text{sinc}\left(\frac{\Delta\beta L}{2}\right)
\end{equation}
Here, $E(\omega)$ is the spectral amplitude of the pump, and the two pump photon frequencies $\omega_{p1}$ and $\omega_{p2}$ are integrated over. It can usually be assumed that the two are approximately equal, then $F$ is just the convolution of $E(\omega)$ with itself, and the delta function, which results from energy conservation, fixes $\omega_p=(\omega_s+\omega_i)/2$. For instance if the pump amplitude is a Gaussian, $E(\omega_p)=E_0e^{-\frac{(\omega_p-\omega_{p0})^2}{2\sigma^2}}$, then
\begin{equation}
F=E_0^2e^{-\frac{(\omega_s+\omega_i-2\omega_{p0})^2}{16\sigma^2}}.
\end{equation}

\begin{figure}
\begin{centering}
\includegraphics[width=0.5\textwidth]{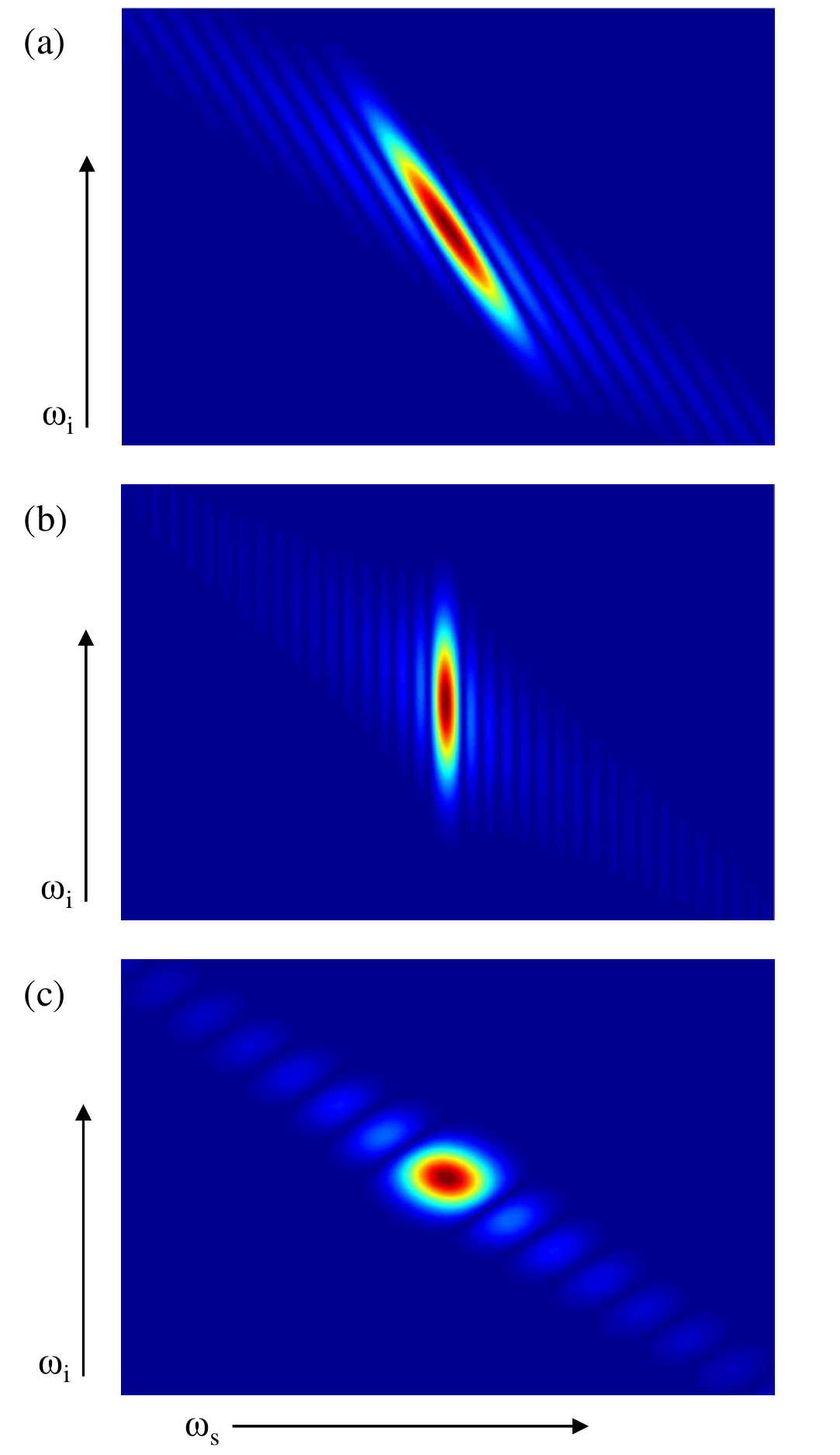}
\caption{Joint spectral amplitudes calculated simply from the pump, signal, and idler group velocities. (a)General example with $\beta_{1p}>\beta_{1i}>\beta_{1s}$ results in a highly correlated signal and idler, indicated by the diagonal nature of the main peak (b) $\beta_{1p}=\beta_{1i}$ results in the central peak becoming vertical and uncorrelated (c) when $\beta_{1p}-\beta_{1s}\approx\beta_{1i}-\beta_{1p}$, the phase-matching condition lies at $+45^\circ$ while the energy matching lies at $-45\%$. Tuning the pump bandwidth can make the central peak circular and uncorrelated. All JSAs are plotted as absolute values, ignoring the complex phase.\label{simpleJSA}}
\end{centering}
\end{figure}
Another useful simplification is to consider the different group velocities at the pump, signal, and idler frequencies but ignore higher-order dispersion terms. Then, for small departures from an exactly phase-matched solution $\Delta\omega_s$ and $\Delta\omega_i$, the phase mismatch can be expressed as:
\begin{equation}
\Delta\beta=\Delta\omega_s(\beta_{1p}-\beta_{1s})+\Delta\omega_i(\beta_{1p}-\beta_{1i})-2\gamma P,
\end{equation}
with $\beta_{1m}=d \beta/d\omega_m=1/v_g$, one over the group velocity at each frequency, $m=p,s,i$. It can be seen that the differences in $1/v_g$ between pump, signal, and idler are an important factor in determining the JSA and its degree of correlation, or its factorability. Fig.~\ref{simpleJSA} shows three JSAs calculated approximately in this fashion. In Fig.~\ref{simpleJSA}(a), the group velocities are chosen arbitrarily with $\beta_{1p}>\beta_{1i}>\beta_{1s}$, resulting in a highly correlated JSA. In Fig.~\ref{simpleJSA}(b), the idler is group velocity matched to the pump, resulting in an uncorrelated JSA, as the main peak now has its axes horizontal and vertical, although the ripples to either side, resulting from the sinc function in $G$, remain correlated. This is the asymmetric scheme to generate a factorable JSA~\cite{halder_purestate}. In Fig.~\ref{simpleJSA}(c), $\beta_{1s}$ and $\beta_{1i}$ are roughly equally spaced above and below $\beta_{1p}$, the symmetric scheme for a factorable JSA~\cite{soller_purestate}. Here, the bandwidth of the pump must be exactly tuned to make the main peak circular in shape and uncorrelated. The conditions on the group velocities are generally met by using birefringent phase-matching, with the pump polarized on the slow birefringent axis and the photons on the fast axis.

For a given JSA, the degree of correlation can be calculated using the singlular value decomposition function in Matlab. This provides a Schmidt decomposition of the JSA:
\begin{equation}
\text{JSA}=\sum_j \lambda_j f_j(\omega_s) g_j(\omega_i),
\end{equation}
where the $f_j(\omega_s)$ and $g_j(\omega_i)$ are a set of orthogonal spectral modes for signal and idler, and the $\lambda_j$ are real amplitude coefficients. We define the purity as
\begin{equation}
P=\sum_j \lambda_j^4,
\end{equation}
which is an upper limit on the quantum interference visibility possible between two photons from separate pairs~\cite{mosley_purity}. For the JSAs in Fig.~\ref{simpleJSA}, the purities are found to be $23\%$ for the correlated JSA in (a), $95\%$ for the asymmetric uncorrelated case in (b), and $83\%$ for the symmetric case in (c). In (c), the purity is mainly limited by the correlation in the sinc function ripples to either side of the main peak. These can be removed by filtering, resulting in high purities with relatively little cost to transmission efficiency~\cite{soller_purestate}.

The nonlinear correction $-2\gamma P$ to $\Delta\beta$, which has been ignored in the calculations above, depends on the pump power. For a pulsed pump, $P$ is a function of position and time, and cannot be simply expressed in the frequency domain- previously the peak power has been used. Instead, we take SPM and XPM into account by working in the time domain, and developing expressions for the joint temporal amplitude (JTA), which is linked to the JSA by 2D fourier transform.

\section{Equations of motion in a nonlinear fibre}

To model photon pair production through FWM with SPM and XPM included exactly, we use the equations of motions with position and time for the pump, signal, and idler fields in a $\chi^{(3)}$ medium. The electric field associated with the pump pulse can be split into positive and negative frequency components as:
\begin{equation}
\textbf{E}=\textbf{\large e\normalsize}X(x,y)\left(E^+_pe^{i(\beta_{p0}z-\omega_{p0}t)}+E^- _pe^{-i(\beta_{p0}z-\omega_{p0}t)}\right).
\end{equation}
Here, $\textbf{\large e\normalsize}$ is the polarization vector of the electric field. $X(x,y)$ is the transverse mode shape, normalised such that $\iint X(x,y)^2 dx dy=1$. Complex oscillatory terms have been separated out at a central frequency $\omega_{p0}$ and wave vector $\beta_{p0}$. This leaves $E^+_p$ as a complex envelope function describing the pulse, dependent on time $t$ and position $z$ along the propagation axis $z$. $E^-_p$ is the complex conjugate of $E^+_p$.

Assuming the pulse envelope varies slowly (the length of the pulse contains many optical cycles, or equivalently the bandwidth of interest is small compared to the frequency $\omega_{p0}$), and that the nonlinearity is a small perturbation to the linear evolution of the pulse, $E^+_p$ obeys the nonlinear Schr\"{o}dinger equation~\cite{agrawal}:
\begin{equation}
\frac{\partial E^+_p}{\partial z}+\beta_{1p}\frac{\partial E^+_p}{\partial t}+\frac{i\beta_{2p}}{2}\frac{\partial^2 E^+_p}{\partial t^2}=i\gamma_p|E^+_p|^2E^+_p.
\label{eqmop}
\end{equation}
The $\beta_{1p}$ term corresponds to the group velocity of the pump pulse. In the following it is removed from the equation as we consider all quantities in a moving reference frame. The $\beta_{2p}$ term gives rise to dispersion - higher order dispersion terms are neglected here, but can be included as higher-order time derivatives. The term on the right-hand side is the nonlinearity associated with SPM of the pump, with $\gamma_p$ an effective nonlinear coefficient:
\begin{equation}
\gamma_m=\frac{3\chi^{(3)}\omega_m}{2cn_mA},
\end{equation}
where $n_m$ is the refractive index at the frequency $\omega_m$ and $A$ the effective area of the transverse mode.

Analagous equations of motion for the signal and idler fields can be written
\begin{equation}
\frac{\partial E^+_s}{\partial z}+\beta_{1s}\frac{\partial E^+_s}{\partial t}+\frac{i\beta_{2s}}{2}\frac{\partial^2 E^+_s}{\partial t^2}=i\gamma_{s}\left(2|E^+_p|^2E^+_s+E^{+2}_pE^-_i\right)
\label{eqmos}
\end{equation}
\begin{equation}
\frac{\partial E^+_i}{\partial z}+\beta_{1i}\frac{\partial E^+_i}{\partial t}+\frac{i\beta_{2i}}{2}\frac{\partial^2 E^+_i}{\partial t^2}=i\gamma_{i}\left(2|E^+_p|^2E^+_i+E^{+2}_pE^-_s\right)
\label{eqmoi}
\end{equation}
In the moving reference frame, $\beta_{1s}$ and $\beta_{1i}$ are taken to be group velocity terms relative to the pump (ie. $\beta_{1s}\rightarrow \beta_{1s}-\beta_{1p}$ and $\beta_{1i}\rightarrow \beta_{1i}-\beta_{1p}$). The first term on the right hand side of these equations is XPM, as the pump modifies the refractive index experienced by signal and idler. The second term relates to FWM, with the strong pump field creating a coupling between signal and idler fields. For simplicity, the central frequencies of the signal and idler have been chosen to be a point of exact phase-matching with the central frequency of the pump, so that:
\begin{equation}
\begin{array}{c}
2\omega_{p0}-\omega_{s0}-\omega_{i0}=0\\
2\beta_{p0}-\beta_{s0}-\beta_{i0}=0.
\end{array}
\end{equation}
Note that since the pump field is many orders of magnitude brighter than the signal and idler, which on average will contain less than one photon, terms representing SPM of signal and idler, and XPM from signal or idler to other fields, are ignored. Similarly depletion of the pump due to FWM is neglected.

The $\chi^{(3)}$ coefficient will generally be three times smaller in nonlinear effects coupling fields of orthogonal polarization compared to fields which are all co-polarized~\cite{agrawal}. In the birefringent phase-matching schemes considered, signal and idler are orthogonally polarized to the pump, so this is incorporated by reducing $\gamma_s$ and $\gamma_i$ by a factor of three, while keeping $\gamma_p$ as above.

Although the pump laser can continue to be treated classically, the signal and idler fields should be quantised. We use the following quantisation, similar to~\cite{huttner_quantise, dot_quantise}:
\begin{equation}
\hat{E}^+_s=\int d\omega \sqrt{\frac{\hbar\omega}{4\pi\epsilon_0cn_\omega}} \hat{a}_\omega e^{-i(\beta_{s0}z+\omega t-\omega_{s0}t)},
\end{equation}
where the $\hat{a}_\omega$ are annihilation operators for a photon at position $z$ with frequency $\omega$, with Hermitian conjugate creation operators $\hat{a}^\dagger_\omega$. An identical expression applies to the idler, with the integral taken to be over a different range of frequencies, so that the two remain distinct. The frequency modes are continuous, so the creation and annihilation operators have a commutation relation $[\hat{a}_\omega,\hat{a}^\dagger_{\omega'}]=\delta(\omega-\omega')$, and the number density operator $\hat{a}^\dagger_\omega \hat{a}_\omega$ should be integrated over a frequency interval in order to refer to the actual number of photons within that interval.


It is convenient to consider the signal and idler in terms of the creation and annihilation operators for a photon at a particular time and position, $\phi^\dagger(z,t)$ and $\phi(z,t)$, which are the fourier transforms (from frequency to time) of $\hat{a}_\omega^\dagger$ and $\hat{a}_\omega$. If the frequencies of interest for signal and idler lie in a narrow bandwidth about $\omega_{s0}$ and $\omega_{i0}$, the electric fields have a simple approximate relation to the new operators:
\begin{equation}
\hat{E}^+_s=\sqrt{\frac{\hbar\omega_{s0}}{2\epsilon_0cn_s}}\phi_s~~~~~~~\hat{E}^+_i=\sqrt{\frac{\hbar\omega_{i0}}{2\epsilon_0cn_i}}\phi_i.
\end{equation}
Like $\hat{E}^+_{s,i}$, the $\phi_{s,i}$ have had the quickly varying oscillations with $z$ removed.

These equations of motions do not have convenient solutions, even the classical equation for the pump, which is independent of signal and idler, unless we neglect the dispersion terms $\beta_2$. Fortunately, as above, the purity largely depends on the different group velocities for the pump, signal, and idler, and the approximate solutions ignoring dispersion are still instructive. To include dispersion properly, numerical methods can be used, as described later.

\section{Approximate solutions}

Once the $\beta_1$ term is removed from equation~\ref{eqmop} using a moving reference frame, and the $\beta_2$ term is ignored, the pump pulse will retain its temporal shape as it propagates down the fibre, only accumulating a nonlinear phase due to SPM:
\begin{equation}
E^+_p(z)=E^+_p(0)e^{i\theta_p}
\end{equation} 
where
\begin{equation}
\theta_p=\gamma_p z |E^+_p(0)|^2
\end{equation}
Below, $E^+_p$ is taken to mean the pump amplitude as a function of time at $z=0$. The signal and idler equations still include a group velocity term, in addition to XPM and FWM terms. In the interaction picture of quantum mechanics, the group velocity and XPM parts of the evolution, which affect signal and idler individually, are applied to the operators, while the FWM interaction between signal and idler is applied to the wavefunction $\ket{\psi}$. We first write the solutions $\phi_{s,i}$ to the group velocity and XPM terms, ignoring FWM:
\begin{equation}
\phi_s(z,t)=\phi_s(0,t-\beta_{1s}z)e^{i\theta_s},
\label{phisolution}
\end{equation}
where the nonlinear phase aquired due to XPM is
\begin{equation}
\theta_s=\frac{2\gamma_s}{\beta_{1s}}\int^{t}_{t-\beta_{1s}z} |E^+_p|^2 dt,
\end{equation}
with an analogous solution for $\phi_i$. Note that because the signal (or idler) experiences group-velocity walk-off from the pump it accumulates phase from the pump at a range of different times, hence the integral. If the signal were group velocity matched to the pump, $\beta_{1s}=0$, the phase would become $\theta_s=2\gamma_sz|E^+_p(t)|^2$.

The evolution of the wavefunction according to FWM is now given by:
\begin{equation}
\frac{d}{dz}\ket{\psi}=i\hat{H}\ket{\psi},
\end{equation}
with
\begin{equation}
\hat{H}=\sqrt{\gamma_s\gamma_i}\int dt E_p^{+2}\phi_s^\dagger\phi_i^\dagger e^{2i\theta_p}+E_p^{-2}\phi_s\phi_ie^{-2i\theta_p}.
\end{equation}
Since the pair rate per pulse will generally be small, to avoid multi-pair emission, we take the interaction to first order, beginning with the signal and idler modes in the vacuum state:
\begin{equation}
\ket{\psi}=\ket{vac}+i\sqrt{\gamma_s\gamma_i}\iint^L_0 dt dz E_p^{+2}\phi_s^\dagger\phi_i^\dagger e^{2i\theta_p} \ket{vac},
\end{equation}
with $L$ the fibre length. To extract the JTA from this wavefunction, we take the overlap between $\ket{\psi}$ and a signal-idler pair at times $t_s$, $t_i$:
\begin{equation}
\text{JTA}(t_s,t_i)=\bra{vac}\phi_s(L,t_s)\phi_i(L,t_i)\ket{\psi},
\label{eq24}
\end{equation}
Substituting in equation~\ref{phisolution} and simplifying we have:
\begin{equation}
\text{JTA}(t_s,t_i)=
\begin{cases}
\frac{i\sqrt{\gamma_s\gamma_i}}{\beta_{1s}-\beta_{1i}}e^{i\Theta}E_p^+(t_c)^2 & \text{if  } 0<z_c<L \\
0 & \text{otherwise}
\end{cases}
\end{equation}
with $z_c$ the point in the fibre at which the pair was created. This is defined for a particular $t_s$,$t_i$ because the signal and idler must be created at the same time, $t_c$, and the extent to which they have walked off from each other identifies the length they have propagated through after creation. Similarly $t_c$ is defined by the differing arrival times of signal and idler:
\begin{equation}
z_c=L-\frac{t_s-t_i}{\beta_{1s}-\beta_{1i}}~~~~~~~~~~~t_c=\frac{\beta_{1s} t_i-\beta_{1i} t_s}{\beta_{1s}-\beta_{1i}}
\end{equation}
$\Theta$ is the nonlinear phase containing the contributions from SPM of the pump up until the pair-creation, and XPM from the point of pair-creation until the end of the fibre:
\begin{equation}
\Theta=2\gamma_pz_c|E_p^+(t_c)|^2 + \frac{2\gamma_s}{\beta_{1s}}\int^{t_s}_{t_c}|E_p^+|^2 dt+ \frac{2\gamma_i}{\beta_{1i}}\int^{t_i}_{t_c}|E_p^+|^2 dt.
\label{NLphase}
\end{equation}

The total probability of generating a pair, or the generation rate per laser pulse, is given by
\begin{equation}
R=\iint |\text{JTA}|^2 dt_s dt_i=\frac{\gamma_s\gamma_iL}{|\beta_{1s}-\beta_{1i}|}\int |E_p^+|^4 dt.
\end{equation}
Clearly this solution becomes unphysical if $\beta_{1s}=\beta_{1i}$, in which case the integrals implicit in equation~\ref{eq24} need to be dealt with differently. Also when dispersion is included it will affect $R$ by causing the pump pulse to broaden or compress in time, respectively decreasing or increasing the generation rate. In the following $R$ is by numerical integration of a JTA over $t_s$ and $t_i$.

\subsection{Asymmetric scheme}

\begin{figure}[b]
\begin{centering}
\includegraphics[width=0.5\textwidth]{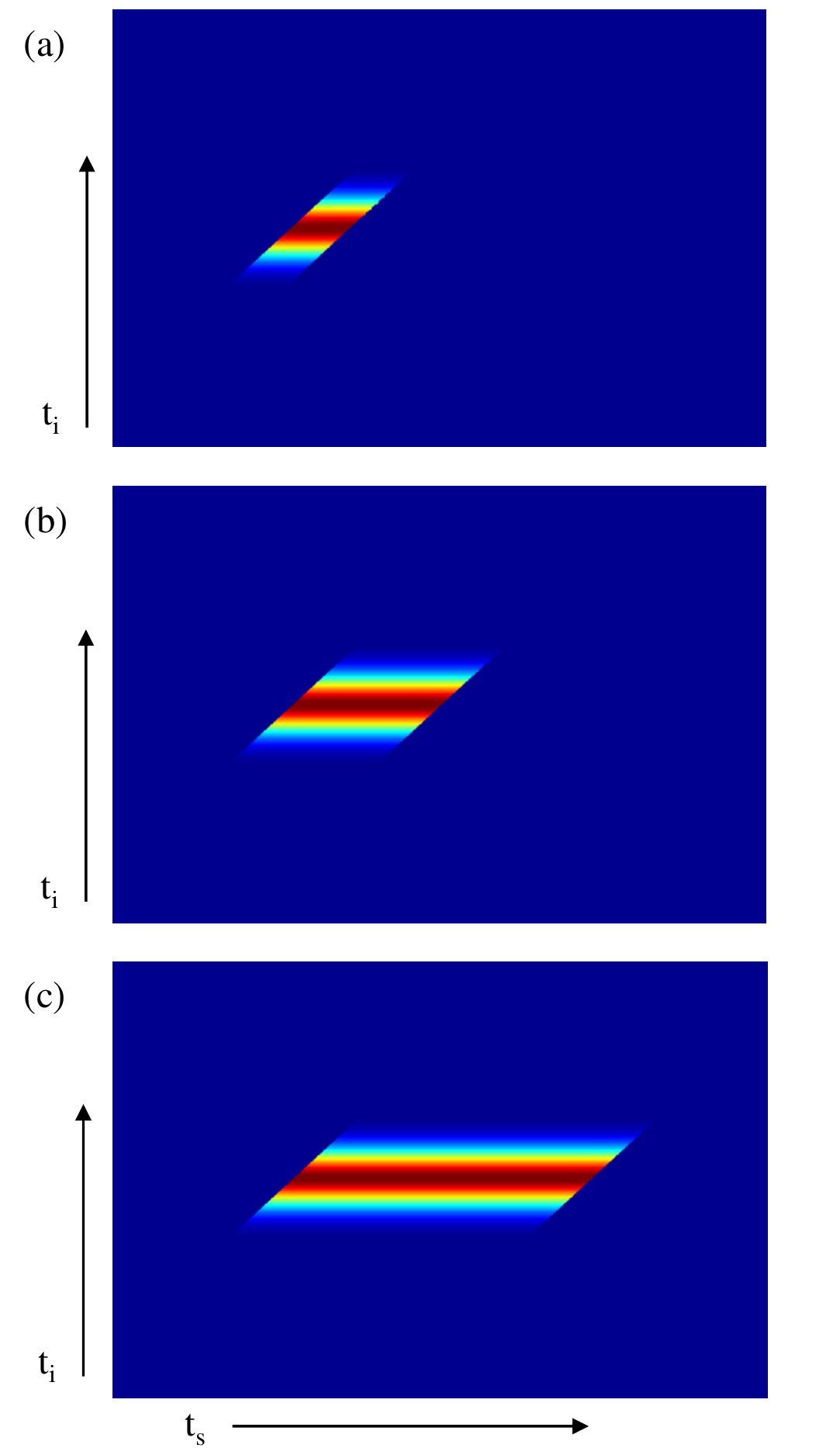}
\caption{Joint temporal amplitudes when the idler and pump are group velocity matched. For increasing fibre length compared to the length of the pump pulse $\tau$, the group velocity walk-off of the signal smears out the JTA. (a) $\beta_{1s}L/\tau=2$ (b) $\beta_{1s}L/\tau=5$ (c) $\beta_{1s}L/\tau=10$.  \label{JTAs}}
\end{centering}
\end{figure}

When the idler and pump are group velocity matched, $\beta_{1i}=0$, $t_c=t_i$. Also the last term in $\Theta$, representing XPM from pump to idler, becomes
\begin{equation}
2\gamma_i(L-z_c)|E_p^+(t_i)|^2,
\end{equation}
because there is no walk-off between pump and idler.

Figure~\ref{JTAs} shows the joint temporal amplitude for three different lengths of fibre, with a Gaussian pump shape $E_p^+\propto e^{-t^2/2\tau^2}$. It can be seen that for short lengths, the photons are highly correlated in time, with hard edges to the JTA due to the condition $0<z_c<L$. As the length is increased, the signal walk-off smears out the JTA, making it closer to rectangular and less correlated. The fibre length is not important in itself so much as the ratio between the fibre length and the length of the pump pulse $\tau$, as changing both by a constant factor is simply a rescaling of the JTA. It will appear uncorrelated if $\beta_{1s}L/\tau\gg 1$, though in practice this may be limited by dispersive effects which have been ignored so far, since they will become more significant for longer lengths and shorter pulses.

\begin{figure}
\begin{centering}
\includegraphics[width=0.5\textwidth]{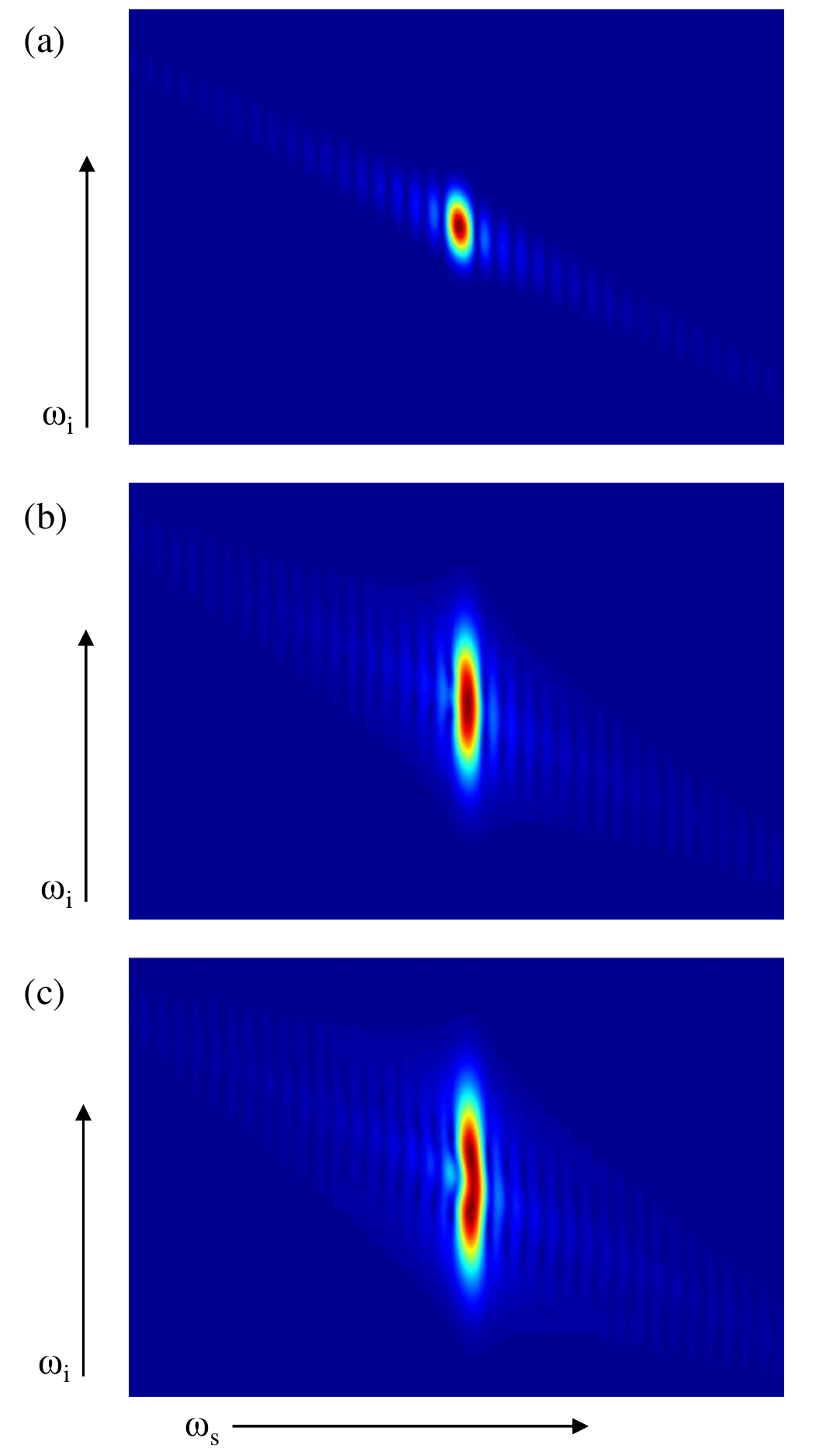}
\caption{JSA for different generation rates $R$, showing the effect of SPM and XPM. The initial effect is to broaden the idler, eventually leading to a splitting and distortion of the JSA which reduces the purity $P$. (a) $R$ approaching zero, $P=89\%$ (b) $R=0.1$, $P=83\%$ (c) $R=0.2$, $P=78\%$.  \label{JSAs_simplePM}}
\end{centering}
\end{figure}

The JTAs are shown as absolute values and so are not affected by the nonlinear phase. The effects can be seen in the JSA obtained by taking the fourier transform of the JTA - in figure~\ref{JSAs_simplePM}, the JSA is shown for increasing probability of pair-creation $R$ in a pulse, in each case with $\beta_{1s}L/\tau=10$. As $R$ increases, the idler is broadened significantly by phase modulation and begins to distort in profile. This causes the purity to decrease, from $89\%$ at $R=0$, to $83\%$ at $R=0.1$, to $78\%$ at $R=0.2$ (0.2 pairs per pulse may be unrealistically high for an experiment, but allows the distortion of the JSA to be seen clearly).

\begin{figure}
\begin{centering}
\includegraphics[width=0.5\textwidth]{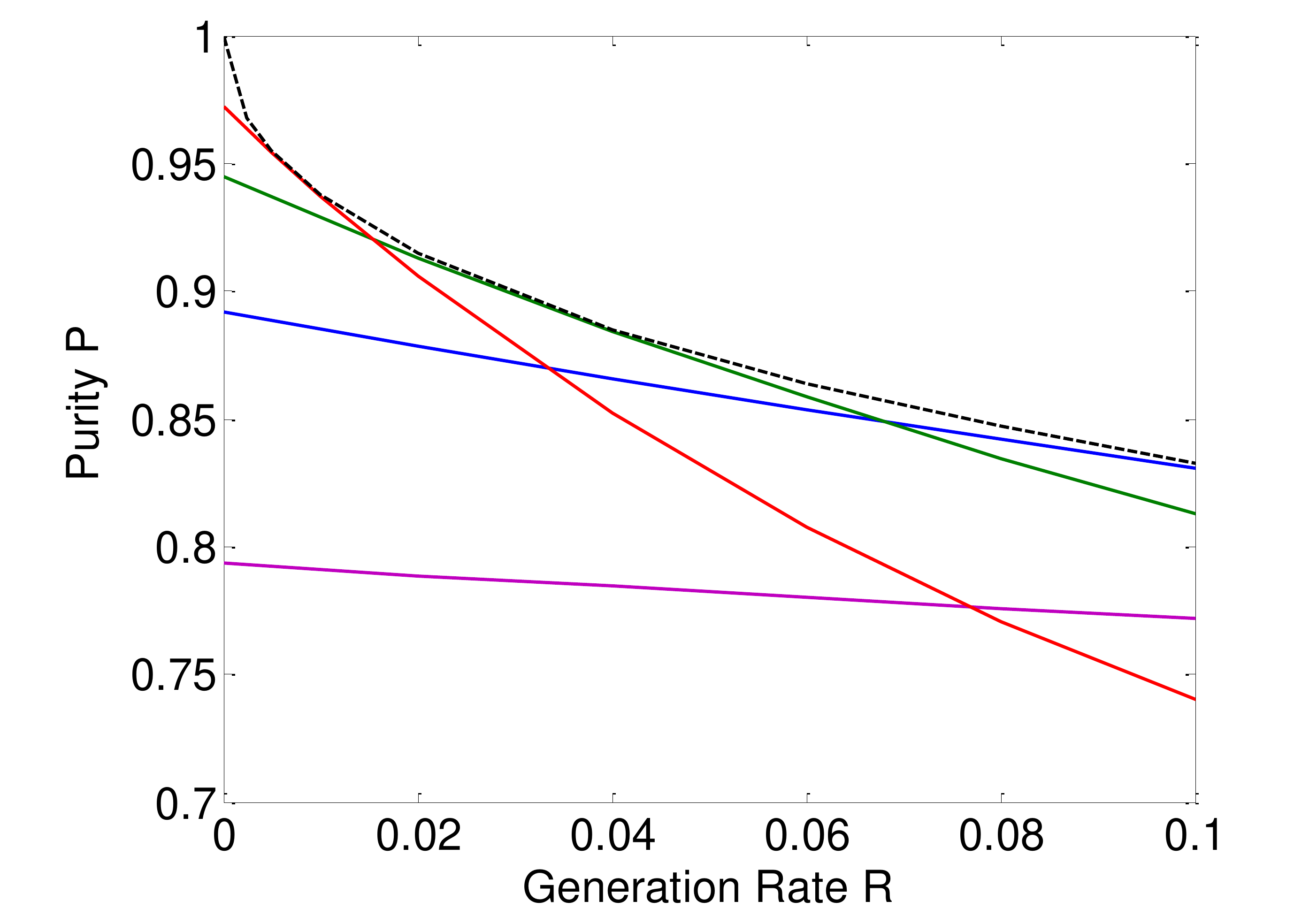}
\caption{Purity plotted against generation rate for four choices of fibre parameters. Red: $\beta_{1s}L/\tau=40$. Green: $\beta_{1s}L/\tau=20$. Blue: $\beta_{1s}L/\tau=10$. Purple: $\beta_{1s}L/\tau=5$. The black dashed line shows the purity varying with generation rate after numerical optimization of  $\beta_{1s}L/\tau$. It can be seen a large value of $\beta_{1s}L/\tau$ is desirable at low rates, but for higher rates it will cause a more rapid fall-off in purity. \label{PurityVsGenRate}}
\end{centering}
\end{figure}

In Figure~\ref{PurityVsGenRate}, the purity is plotted against the pair generation rate $R$ for different choices of fibre parameters. It can be seen that for larger values of $\beta_{1s}L/\tau$, the purity will be high at very low $R$, but will decrease rapidly as $R$ increases, whereas a smaller value of $\beta_{1s}L/\tau$ will decrease more slowly and may be optimal for a given $R$. Note that, even after optimising the purity with the fibre parameters, the purity will be more detrimental to quantum interference quality than multi-pair emission over the range shown, $0< R\leq 0.1$. [The probability of multi-pair emission is estimated as $R^2$, which when compared to the rate of single-pair emission $R$ can reduce the interference visibility by at most $R$]. This is potentially significant if the end goal is to build a deterministic photon source by multiplexing together many of these sources~\cite{ma_multiplex}, then to achieve high-quality interference without filtering for quantum communication or computing applications. It is usually assumed that the end quality will be high so long as multi-pair emission is kept low from the individual sources, but this shows that, at least for this asymmetric scheme using FWM, the effects of phase modulation are likely to be the limiting factor on the generation rate.

Inspection of equation~\ref{NLphase} does suggest a solution to this problem. The nonlinear phase factor $e^{i\Theta}$ becomes a factorable function of $t_s$ and $t_i$ over the extent of the JTA so long as the pump field $E_p^+$ is a square function in time. The phase is only correlated because $|E_p^+(t_c)|^2$ is varying across the JTA. However, the effects of group-velocity dispersion acting on a short, square pulse over a large length may be unpleasant. Using realistic dispersion parameters based on the birefringent microstructured fibre in~~\cite{tame_simons1, bell_secretsharing1} with a length $50cm$, and calculating the JSA from equation~\ref{JSA_FxG}, without phase modulation effects, the maximum value of $P$ using a square pulse of optimal duration is found to be $80\%$. Prechirping the pulse to compensate the dispersion, so that it is square at the midpoint of the fibre, yields a slight improvement to $81\%$. So this is unlikely to be helpful unless the dispersion is particularly small.

Another solution would be to have $\gamma_p=\gamma_i$, although this is not possible using the birefringent phase-matching considered, because of the reduction in the effective nonlinearity by a factor of 3 when the fields are orthogonally polarized. However if the nonlinear phase could be made factorable, a high purity could again be achieved with a large value of  $\beta_{1s}L/\tau$. 

\begin{figure}[h!]
\begin{centering}
\includegraphics[width=0.5\textwidth]{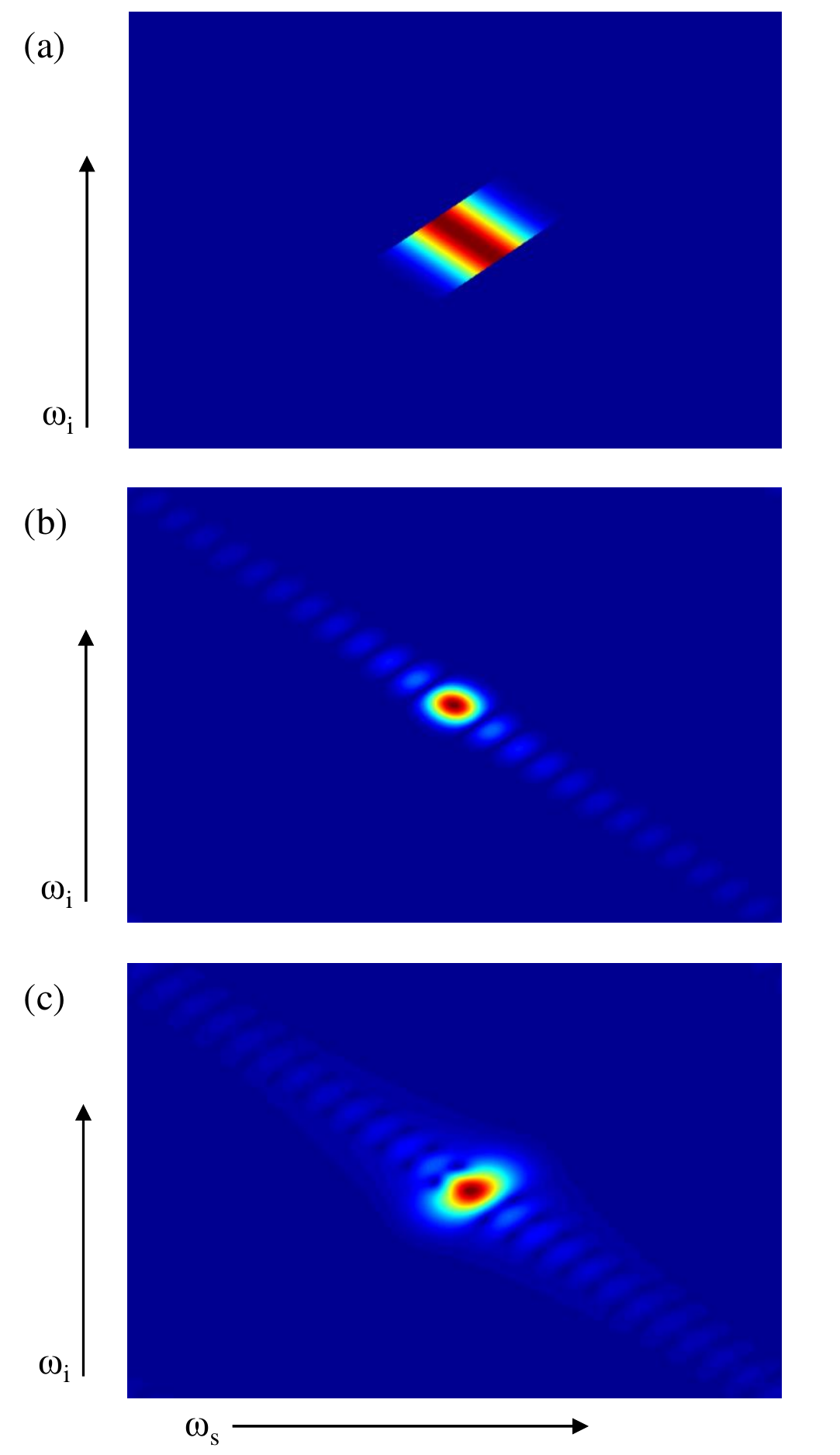}
\caption{(a) Joint temporal amplitude when the signal and idler are equally spaced in $\beta_1$ about the pump. The pump duration is optimised to avoid correlation, but the hard edges to the JTA caused by the sudden beginning and end of the nonlinearity mean that some correlation is inevitable. (b) Corresponding JSA with low generation rate. (c) At higher generation rate, $R=0.2$, showing the effects of SPM and XPM. The JSA is broadened in one direction. \label{SymmAmps}}
\end{centering}
\end{figure}

\subsection{Symmetric scheme}

\begin{figure}[h!]
\begin{centering}
\includegraphics[width=0.5\textwidth]{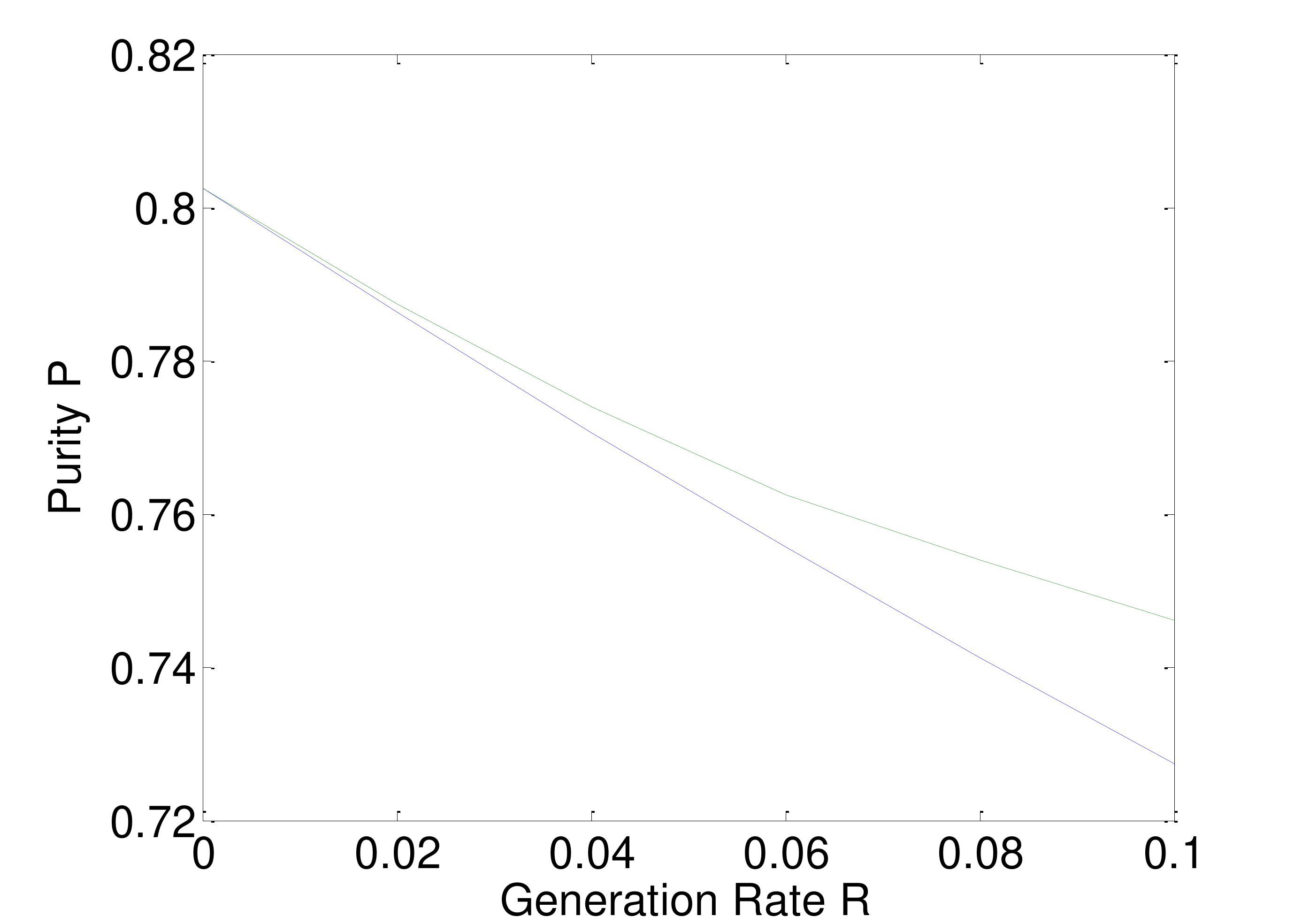}
\caption{Purity against generation rate for the symmetric scheme. Blue: $\tau$ is kept constant. Green: $\tau$ is increased with $R$ to reoptimise. Again, the phase modulation reduces $P$ as $R$ is increased, though here $P$ is lower to start with than in the asymmetric scheme because there is more correlation in the sinc ripples of the JSA. \label{PurityVsGenRateSymm}}
\end{centering}
\end{figure}

We now consider the symmetric scheme for avoiding correlations, with $\beta_{1s}=-\beta_{1i}=\beta_1$. This implies that $z_c=L-\frac{t_s-t_i}{2\beta_1}$ and $t_c=\frac{t_s+t_i}{2}$. Figure~\ref{SymmAmps}(a) shows the JTA in this case, where the temporal width $\tau$ of a gaussian pump has been optimised to minimise correlation. Figure~\ref{SymmAmps}(b) shows the corresponding JSA without the effects of phase modulation, with $R$ approaching zero, while Figure~\ref{SymmAmps}(c) shows the broadening and distortion from phase modulation when $R=0.2$. Here, the broadening introduces spectral correlation, but it can be partly compensated by beginning with a longer pump pulse (increasing $\tau$). Figure~\ref{PurityVsGenRateSymm} shows the purity plotted against generation rate, both for a fixed value of $\beta_1L/\tau$, and with $\tau$ reoptimised as $R$ is increased.

\begin{figure}[h!]
\begin{centering}
\includegraphics[width=0.5\textwidth]{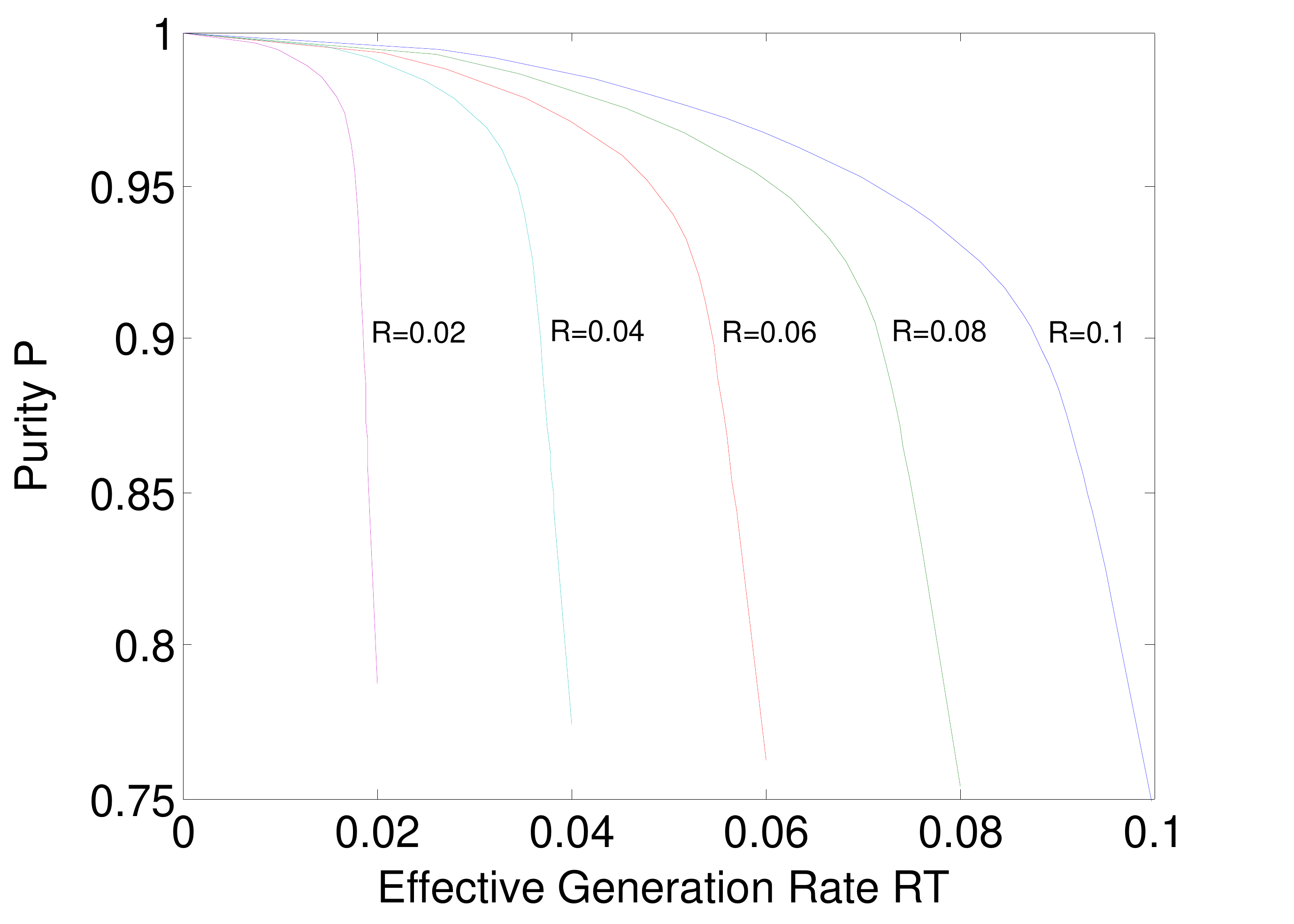}
\caption{Purity as a function of the effective generation rate, $RT$, when filtering is applied to one of the photons (the herald) resulting in transmission $T$. Unfiltered generation rates $R$ are shown from 0.02 to 0.1.  \label{FilteredPurity}}
\end{centering}
\end{figure}

The predicted purities of around $80\%$ here are somewhat low, even when $R$ is kept small. A realistic experimental strategy may be to increase the purity above this by applying some spectral filtering to one photon of the pair, with an overall transmission $T$. If the filtered photons are used as the heralds, the heralded photons will not experience any loss, just a reduction in effective generation rate to $RT$. Figure~\ref{FilteredPurity} shows the purity after one of the photons has been filtered with a top-hat transmission window of variable width, as a function of $RT$. Five different values for the original generation $R$ are shown, from 0.2 to 0.1. As expected the purity tends to 1 as the filtering becomes very drastic and only leaves one possible spectrum for the heralded photon, although this restricts the source to low effective generation rates. It can also be seen that the purity tends to 1 faster if the initial (unfiltered) generation rate $R$ is larger, in spite of the detrimental effects of phase modulation.

Since a larger $R$ appears to be beneficial for the purity after filtering, as a function of $RT$, this suggests there will be a trade-off between achieving higher purity and keeping multi-pair emission low, which occurs with probability approximately $R^2$~\cite{mosley_purity}. If photon-number resolving detectors become available, they could be used on the heralds to detect and filter out multi-pairs. Otherwise it may not be possible to simulataneously achieve a high purity, high effective generation rate, and low multi-pair emission using this scheme. In future work, it would be useful to consider similar schemes where two pump frequencies are used, with mismatched group velocities~\cite{fang_2pump}. This can in principle remove the sinc ripples from the JSA, and hence most of the correlation, although phase modulation may still have an effect.

\section{Numerical Model}

To include the effects of group-velocity dispersion accurately, it is necessary to go to a numerical model involving finite-steps along the fibre length. A common method for modelling the propagation of a laser pulse through a nonlinear and dispersive medium is a split-step fourier (SSF) simulation~\cite{agrawal}. Here, the length is divided into small steps $\Delta z$, and for each step, the nonlinearity and the dispersion are applied separately. For instance, the effect of propagating through the nonlinearity of $\Delta z$ could be applied first, in the time domain where this is a simple calculation, then the pulse could be fourier transformed to the frequency domain, where the effect of the dispersion can easily be applied using $\tilde{E}^+(z,\omega)=\tilde{E}^+(0,\omega)e^{ik(\omega)z}$, followed by inverse fourier transform back to the time domain. Since the effects of the nonlinearity and dispersion are generally non-commuting, this is only approximate, but is accurate for small $\Delta z$. In fact, it is better to apply half a step of dispersion, then a full step of nonlinearity, followed by the other half step of dispersion, as then the size of the errors due to the approximation vary with $\Delta z^3$ rather than $\Delta z^2$~\cite{agrawal}.

To model the pair-production process along similar lines, we use a SSF simulation for the propagation of the pump pulse, and incorporate spontaneous FWM into the nonlinear part of each step. The steps in position are kept small compared to the resolution in time, so that $\beta_{1s,i}\Delta z<\Delta t$. This means the signal and idler are intially in the same time-bin as each other, and as the component of the pump which created them, which simplifies the calculation for each step. To propagate the state of pairs created in previous steps, the half-step of dispersion is applied to the JSA, which is then converted to a JTA by 2D fourier transform so that XPM can be applied, before it is transformed back to a JSA for the next half-step of dispersion. The state of the new pairs is coherently added for each step.

In the previous sections, the purity was determined by two parameters: the group velocities relative to the fibre length and pulse duration, $\beta_1 L/\tau$, and, when nonlinear phase modulation was taken into account, the total probability of pair-generation, $R$. Here, the group-velocity dispersions at each wavelength introduce additional relevant parameters: $\beta_2 L/\tau^2$ for pump, signal, and idler. If the fibre length and pulse duration are increased in proportion the effect of group-velocity dispersion decreases.

\begin{figure}
\begin{centering}
\includegraphics[width=0.5\textwidth]{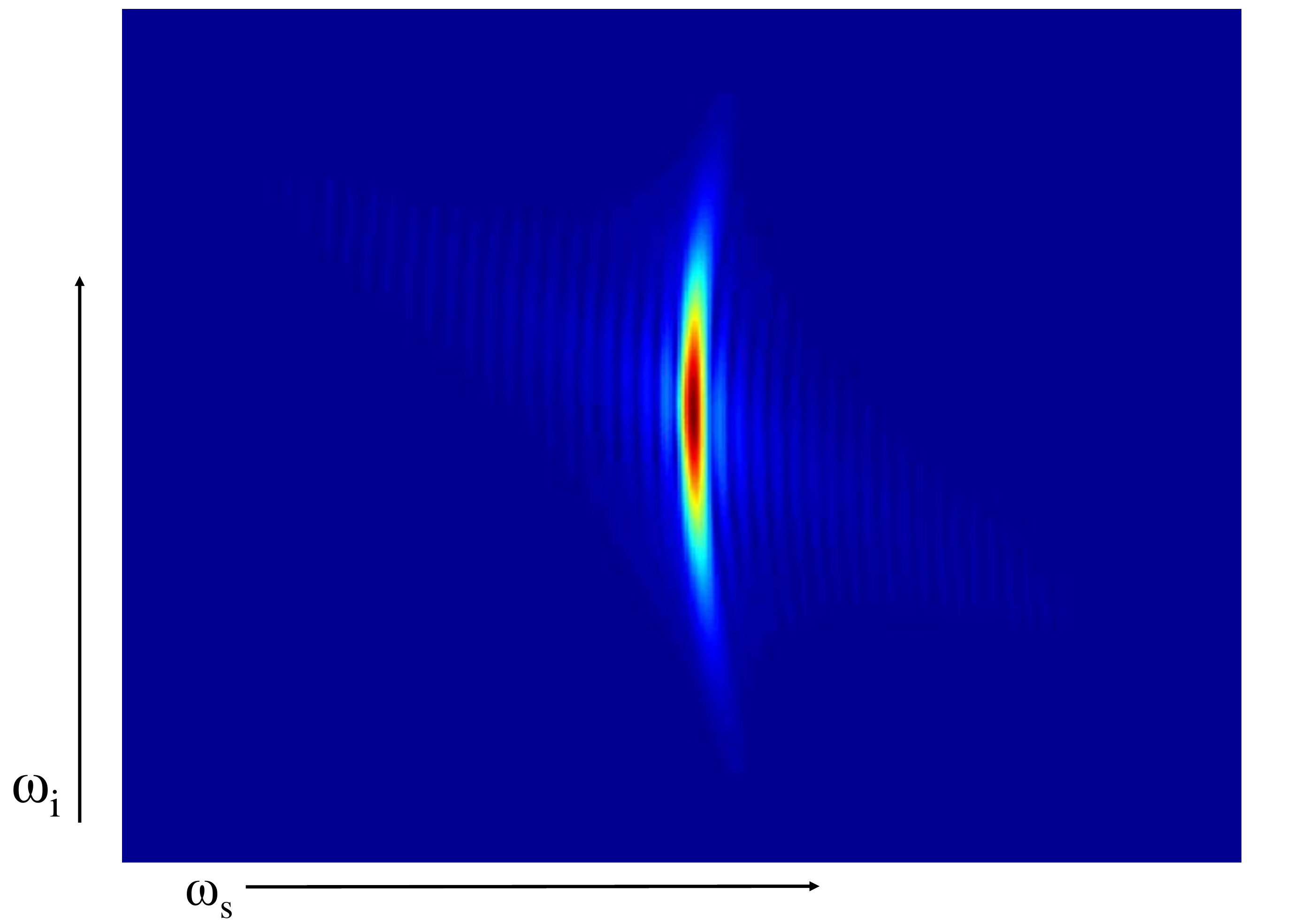}
\caption{JSA from the numerical model including dispersion, for realistic fibre parameters with a length of $50cm$ and a pump bandwidth $2nm$. The dispersion causes some curvature of the JSA, which may introduce correlation. \label{JSAdispersion}}
\end{centering}
\end{figure}

\begin{figure}[b]
\begin{centering}
\includegraphics[width=0.5\textwidth]{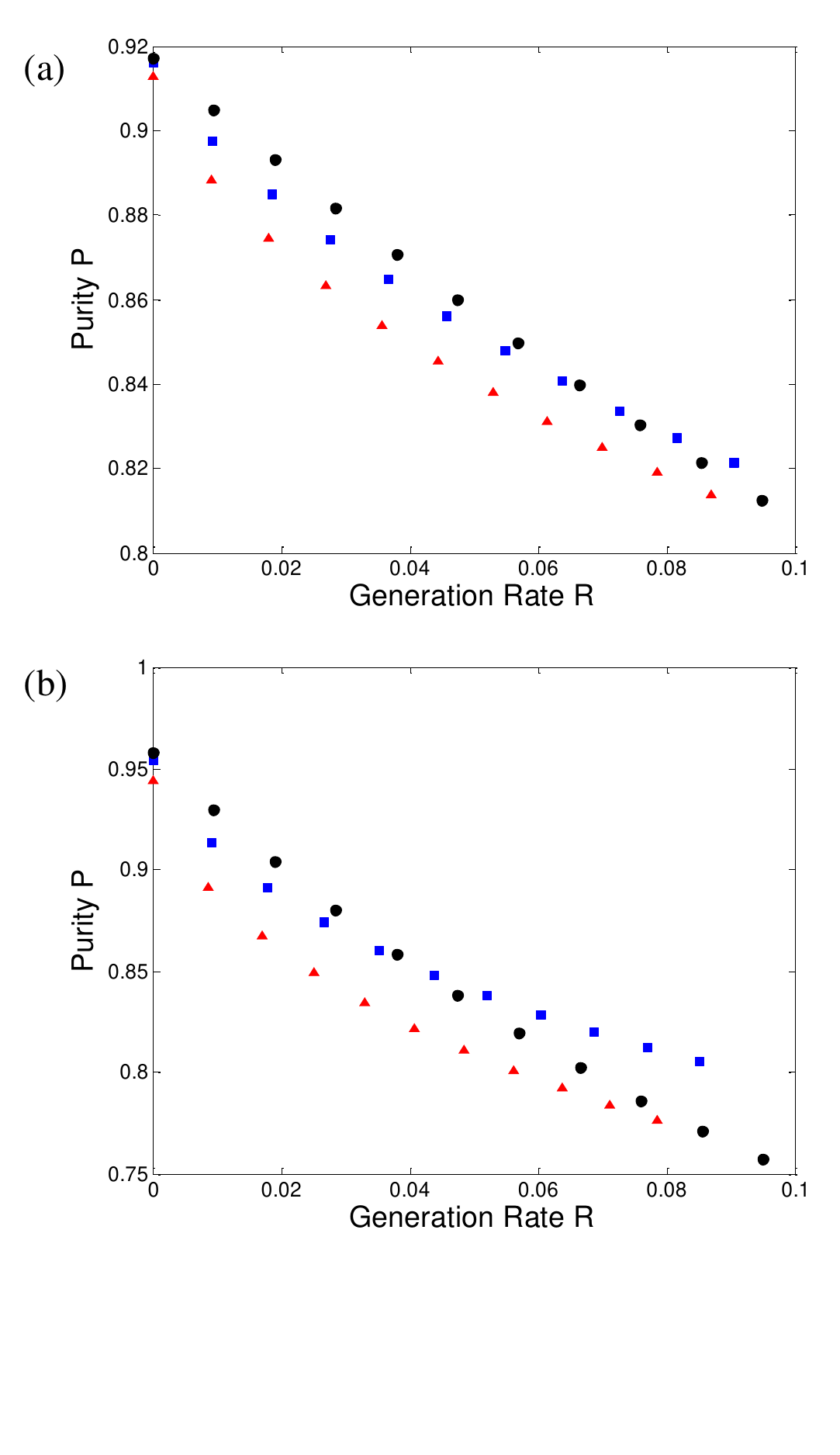}
\vspace{-2.4cm}
\caption{Purity against generation rate $R$ for different strengths of dispersion, with an initial pump bandwidth of $1nm$. $R=0$ corresponds to the case with no non-linear phase modulation. (a) $L=0.5m$,(b) $L=1m$. In each case, black circles: no dispersion; blue squares: realistic dispersion; red triangles: double strength dispersion. Surprisingly, at higher rates the dispersive case sometimes does better than the case with no dispersion.     \label{PvsRnumerical}}
\end{centering}
\end{figure}

We again consider dispersion parameters taken from the birefringent microstructured fibre used in~\cite{tame_simons1, bell_secretsharing1}. This fibre makes use of the asymmetric scheme to avoid correlations, with the pump pulse at $726nm$ group-velocity matched to the idler at $864nm$. The signal, phase-matched at $626nm$, experiences walk-off with $\beta_{1s}=1.14\times 10^{-11}m^{-1}s$. Pump, signal, and idler experience group-velocity dispersion with $\beta_{2p}=2.1\times 10^{-26}m^{-1}s^2$, $\beta_{2s}=3.6\times 10^{-26}m^{-1}s^2$, and $\beta_{2i}=-1.3\times 10^{-26}m^{-1}s^2$. Figure~\ref{JSAdispersion} shows the JSA produced from a $50cm$ fibre with a $2nm$ initial pump bandwidth, corresponding to $\tau\approx 230fs$, and with the generation probability $R=0.05$. The dispersion introduces some curvature to the JSA, so that it will become correlated for large bandwidth pulses.

In figure~\ref{PvsRnumerical}, the purity is plotted against the generation probability using the numerical model, for different amounts of dispersion: no dispersion, dispersion using the realistic parameters, and double strength dispersion. Physically, doubling the strength of the dispersion while keeping the other parameters constant could be achieved by halving the length of the fibre, halving the laser pulse duration, and adjusting the laser power to keep $R$ constant. In figure~\ref{PvsRnumerical}(a), $\beta_{1s} L/\tau$ is approximately 12. It can be seen that for a very low generation rate, where phase modulation is negligible, the dispersion only causes a slight reduction in purity. The zero $R$ intercept is close to our previous modelling results which ignore non-linear phase modulation~\cite{clark_purestate}. However, as $R$ is increased, dispersion has a larger effect, suggesting that dispersion and phase modulation are combining to create more of a reduction in purity than either would alone. Surprisingly, for higher values of $R$, this trend reverses, and the case with some dispersion actually does better than no dispersion. This can be seen more prominently in figure~\ref{PvsRnumerical}(b), where the fibre length was doubled to 1m, so that $\beta_{1s} L/\tau\approx 24$. It seems that for some choices of parameters, the nonlinear phase modulation and the dispersion begin to compensate one another, although here the effect is too small to change the trends seen before, with the purity still decreasing as the generation rate is increased, and with a larger value of $\beta_{1s} L/\tau$ creating a better purity at low rate, but a worse purity at high rates.

\begin{figure}[t]
\begin{centering}
\includegraphics[width=0.5\textwidth]{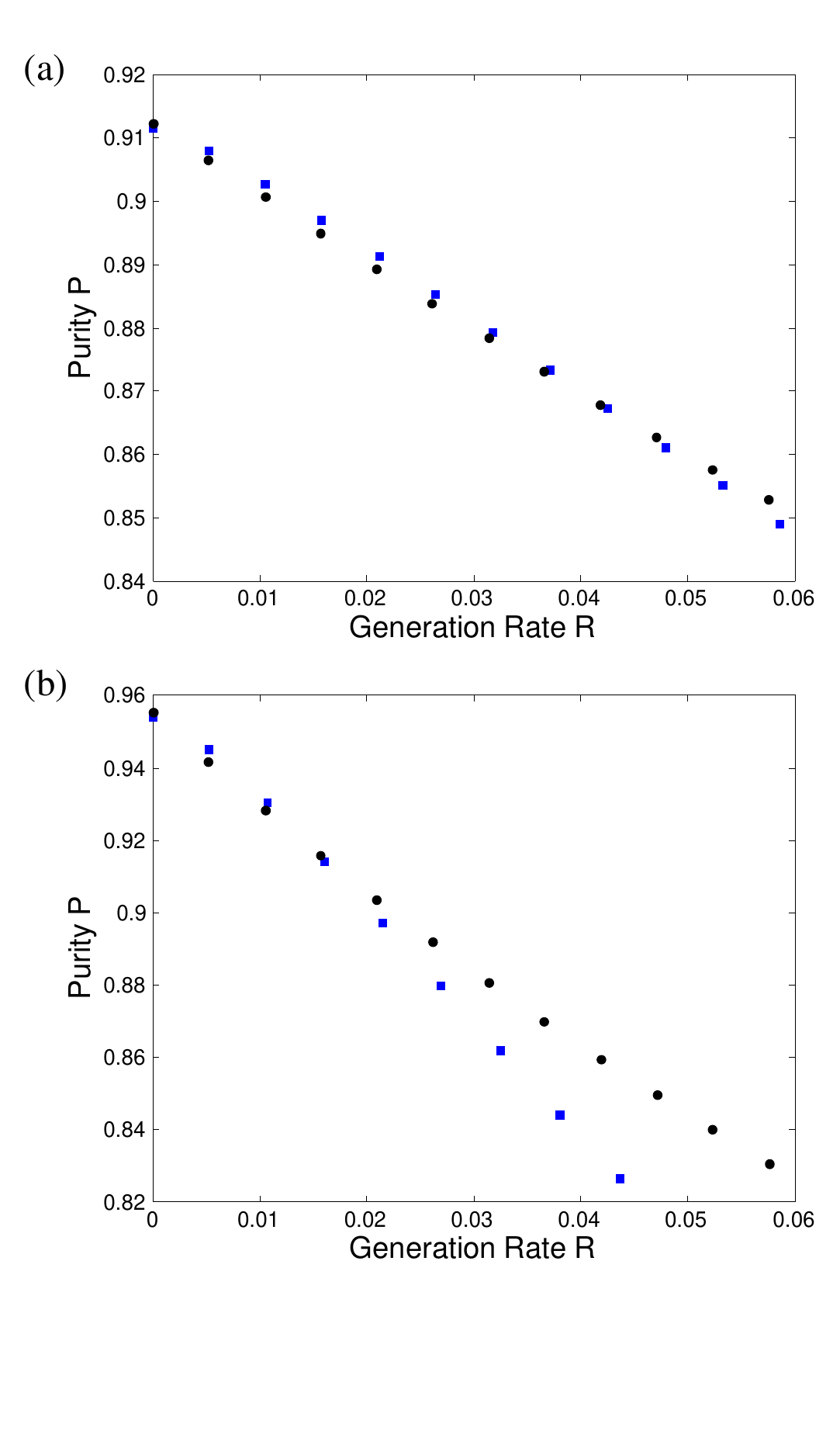}
\vspace{-2.4cm}
\caption{Purity against generation rate with anomalous dispersion, for an initial pump bandwidth of $2nm$. (a) $L=0.5m$,(b) $L=1m$. In each case, black circles: no dispersion; blue squares: realistic dispersion. Here, a very slight benefit can be seen at low rate for the dispersive case compared to the case without dispersion. \label{PvsRnumerical_anomalous}}
\end{centering}
\end{figure}

Finally, we consider a different set of fibre parameters, corresponding to a birefringent microstructured fibre pumped at $1064nm$, in its anomalous dispersion region, generating phase matched photons at $810nm$ and $1550nm$~\cite{mcmillan_source}. The pump is polarized on the fast axis of the fibre while signal and idler are polarized on the slow axis, such that the signal is now group velocity matched to the pump, while the idler walks off with $\beta_{1i}=1.2\times 10^{-11}m^{-1}s$. The dispersion parameters are $\beta_{2p}=-8.7\times 10^{-27}m^{-1}s^2$, $\beta_{2s}=1.0\times 10^{-26}m^{-1}s^2$, and $\beta_{2i}=-6.4\times 10^{-26}m^{-1}s^2$. Figure~\ref{PvsRnumerical_anomalous}(a) shows the purity plotted against generation rate for a $50cm$ length of this fibre with an initial pump bandwidth of $2nm$, resulting in $\beta_{1s} L/\tau\approx 12$, and figure~\ref{PvsRnumerical_anomalous}(b) shows the case when the length is increased to $1m$, so $\beta_{1s} L/\tau\approx 24$. It can be seen that at low $R$, the dispersion improves the purity slightly compared to the case without dispersion but with phase modulation. However, for larger generation rates the combination of dispersion and phase modulation has a significant deleterious effect, as can be seen clearly in figure~\ref{PvsRnumerical_anomalous}(b).

\section{Conclusion}

We have seen that for schemes seeking to minimise the correlation between photon pairs generated by four-wave mixing, the effects of self-phase modulation and cross-phase modulation may be a limiting factor on the photons' purity which is not usually considered. For the asymmetric scheme, where one of the generated photons is group velocity matched to the pump pulse, it would otherwise be expected that, with a long interaction length and a wide pump bandwidth, very high purity could be achieved. However, when these additional nonlinear effects are included, the purity is degraded as the generation rate is increased, which may limit sources to low rates when a particular purity or quantum interference visibility is required. This can be seen both in an analytical model in the time domain, where group-velocity dispersion is neglected, and in a numerical model which includes both nonlinear effects and dispersion.

The symmetric scheme to generate pure photons is also considered, with the signal and idler group indices equally spaced above and below the pump group index. Again the nonlinear effects significantly degrade the purity as the generation rate is increased, although here the main source of impurity is the correlation in the sinc-ripples of the phase-matching function. These ripples can be eliminated with narrow filtering, but the purity only tends to unity as the transmission through the filter becomes small. In future work it would be interesting to consider the case with two pump fields at different wavelengths, where the sinc ripples can in theory be eliminated without filtering, but it seems likely that nonlinear effects will also be detrimental there.

The numerical model demonstrates that the impurity from nonlinear effects and from dispersion do not combine trivially, sometimes leaving a lower purity than would be expected when the effects are taken individually, but in some regimes slightly higher. It is possible that for particular choices of pump pulse power, duration, and shape, the effects of dispersion and nonlinearity could be made to cancel in a soliton-like manner, and leave a high purity, although it is not expected that having the pump pulse alone propagating as a soliton would achieve this.

We conclude that the discrepancy between previous modelling results and measured visibilities in heralded photon interference experiments~\cite{halder_purestate, clark_purestate} can be largely explained by including non-linear phase modulation. We also note that this effect limits the fidelity of cluster states generated by fusing entangled states \cite{tame_simons1, bell_secretsharing1} and thus could limit the scalability of cluster state quantum computation based on four wave mixing

The authors acknowledge support from EU project 600838 QWAD and ERC advanced grant 247462 QUOWSS.

\end{document}